\documentclass[aps,showkeys,preprint,nofootinbib,tightenlines]{revtex4}
\usepackage[a4paper, margin=0.90in]{geometry}
\usepackage{graphicx}  %
\usepackage{caption}
\usepackage{subcaption}
\usepackage{amsmath}
\usepackage{bm}  %
\usepackage{ulem}
\usepackage{mwe}
\usepackage{comment}
\usepackage{verbatim}
\usepackage[titletoc,title]{appendix}
\usepackage{comment}
\usepackage{wrapfig}
\usepackage{lipsum}
\newcommand{\bea}{\begin{eqnarray}}
\newcommand{\eea}{\end{eqnarray}}
\newcommand{\beq}{\begin{equation}}
\newcommand{\eeq}{\end{equation}}
\newcommand{\bqa}{\begin{eqnarray}}
\newcommand{\eqa}{\end{eqnarray}}

\def\mqo2{{\!\!\!}}
\DeclareMathOperator\arctanh{arctanh}

\begin{document}
\title{
Conformality Lost in Efimov Physics}
\author{Abhishek Mohapatra}
\email{mohapatra.16@buckeyemail.osu.edu}
\affiliation{Department of Physics,
         The Ohio State University, Columbus, OH\ 43210, USA\\}

\author{Eric Braaten}
\email{braaten@mps.ohio-state.edu}
\affiliation{Department of Physics,
         The Ohio State University, Columbus, OH\ 43210, USA\\}

\date{\today}
\begin{abstract}
A general mechanism for the loss of conformal invariance is the merger and disappearance of an infrared fixed point
and an ultraviolet fixed 
point of a renormalization group flow.   
We show explicitly how this mechanism works in the case of identical bosons at unitarity as the spatial 
dimension $d$ is varied. For $d$ between the critical dimensions $d_{\rm 1}=2.30$ and $d_{\rm 2}=3.76$, there 
is loss of conformality as evidenced by the Efimov effect in the three-body
sector. The beta function for an appropriate three-body coupling is a quadratic polynomial in that coupling. 
For $d<d_{\rm 1}$ and for $d>d_{\rm 2}$, the beta function has two real roots that correspond to infrared and 
ultraviolet fixed points. As $d$ approaches $d_{\rm 1}$ from below and as $d$ approaches $d_{\rm 2}$ from above, the 
fixed points merge and disappear into the complex plane.
For $d_{\rm 1}<d<d_{\rm 2}$, the beta function has complex roots and the renormalization group flow for the three-body 
coupling is a limit cycle.

\end{abstract}

\smallskip
\maketitle

\section{Introduction}
In a classic paper in 1970 \cite{Efimov:1970}, Vitaly Efimov discovered a remarkable phenomenon
that can occur in a three-body system of nonrelativistic particles with 
short-range interactions if at least two of the three pairs of particles have  two-body scattering 
length $a$ much larger than their range. In the unitary limit $a\rightarrow\pm\infty$, there is 
a shallow two-body bound state exactly at the two-body threshold. 
Efimov showed that the three-body system in the unitary limit can have infinitely many shallow three-body bound 
states with an accumulation point at the three-body threshold. 
This phenomenon is called the {\it Efimov effect}, and the three-body bound states are called {\it Efimov states}.
As the threshold is approached, 
the ratio of the binding energies of successive Efimov states
approaches a universal constant that depends on 
the statistics and the mass ratios of the particles.
This spectrum is evidence for discrete scale invariance in the three-body system. 

The renormalization group (RG) has provided deep insights into  
quantum field theories in condensed matter, particle, and nuclear  physics \cite{Wilson:1983}.
An RG flow describes how parameters that describe short-distance scales 
depend on the ultraviolet (UV) cutoff $\Lambda$. 
The simplest attractor for an RG flow  is a fixed point. 
At a fixed point, the short-distance parameters are independent of the UV cutoff $\Lambda$, 
and  this requires the system to be scale invariant. 
Another possible attractor for an RG flow is an RG \textit{limit cycle},
a possibility first proposed by Wilson \cite{Wilson:1971}. 
The RG flow at a limit cycle is around a closed loop. The RG flow completes one cycle every time 
the UV cutoff $\Lambda$ is changed by a specific multiplicative factor $\lambda_{\rm 0}$. 
This requires the system to have discrete scale invariance with discrete  scaling factor $\lambda_{\rm 0}$.
The discrete scale invariance associated with the Efimov effect can be interpreted in terms of the RG flow
of  a three-body coupling with a limit cycle \cite{Eric_Hammer}.

Scale invariance of a system may be compatible with a larger symmetry group of conformal transformations.
For Lorentz-invariant quantum field theories in two space-time dimensions, 
Zomolodchikov has proved under  general assumptions that scale invariance implies conformal invariance 
\cite{Zomolodchikov:1986}. For relativistic quantum field theories with space-time dimension greater than two, 
there are no known counterexamples to the conjecture
that scale invariance implies conformal invariance \cite{Nakayama:2015}. 
In nonrelativistic quantum field theories with Galilean invariance \cite{Hagen:1972, Niederer:1972},
the conjecture that scale invariance implies conformal invariance
has not been proven even in two space dimensions,  but again
there are no known counterexamples \cite{Nakayama:2015}. 
In this article, we will use the phrases ``scale invariance'' and ``conformal invariance'' interchangeably. 

In a paper entitled ``Conformality Lost" \cite{Kaplan:2009}, Kaplan, Lee, Son, and Stephanov revealed a general 
mechanism for the loss of conformal invariance in a system. As a 
parameter $\alpha$ of a scale-invariant system approaches a critical value $\alpha_*$ where conformality is 
lost, an ultraviolet (UV) fixed point and an 
infrared (IR) fixed point of an appropriate RG flow merge together and disappear. The RG flow
 near the point where conformality is lost can be described by a simple model with a single dimensionless coupling
$g$ whose beta function has the form
\begin{equation}
\Lambda\frac{d~}{d\Lambda}g=\left(\alpha-\alpha_*\right)-\left(g-g_*\right)^2,
\label{model_running}
\end{equation}
where $\Lambda$ is the renormalization scale and $g_*$ is a constant independent of $g$ and $\alpha$. For 
$\alpha>\alpha_*$, the coupling $g$ has a UV fixed point $g_{+}$ and an  IR fixed point $g_{-}$, 
which are given by
\begin{equation}
g_{\pm}=g_{*}\pm\left(\alpha-\alpha_*\right)^{1/2}.
\label{fixedpoints_model}
\end{equation}
As $\alpha$ decreases toward $\alpha_*$, the two fixed points both approach $g_*$,
and they  merge together at $\alpha=\alpha_*$.
For $\alpha<\alpha_*$, the zeros of the beta function
have disappeared into the complex plane and conformality is lost. 

In Ref.~\cite{Kaplan:2009}, the authors suggested that the loss of conformality associated with Efimov physics can be 
understood  in terms of the merging of a UV fixed point and an IR  fixed point for a three-body 
coupling, but they didn't show this explicitly.
In the case of  identical bosons in three spatial dimensions at unitarity $\left(a\rightarrow\pm\infty\right)$, \
the authors pointed out that there is no tunable parameter analogous to $\alpha$
in Eq.~\eqref{model_running}. 
In this paper, we show explicitly how conformality is lost in the case of identical bosons at unitarity when the spatial 
dimension $d$ is the external parameter.  We show that the beta function for an appropriate 
three-body coupling is a quadratic polynomial analogous to Eq.~\eqref{model_running}
with coefficients that depend on $d$. 
There are two critical dimensions $d_{\rm 1}=2.30$ and $d_{\rm 2}=3.76$ at which there is loss of conformality. 
For $d<d_{\rm 1}$ and $d>d_{\rm 2}$, the beta function has real zeros that correspond to a UV fixed
point and an IR fixed point.
As $d$ enters the interval $d_{\rm 1}<d<d_{\rm 2}$ where conformality is lost, 
the zeros of the beta function disappear into the complex plane, 
and the RG flow for the three-body coupling becomes a limit cycle. 
The continuous scaling symmetry is broken down to a discrete 
scaling symmetry that manifests itself in the Efimov effect. 

This article has two additional sections followed by a summary. 
In Section~\ref{two_body}, we consider briefly the two-body sector for identical 
bosons in $d$ dimensions. We determine the beta function for the two-body coupling.  
It has a zero  that corresponds to a  UV fixed point associated with conformal invariance in the 
two-body sector. In Section~\ref{three_body_Efimov}, we 
consider in detail the three-body sector for identical bosons in $d$ dimensions. 
We show that the beta function for a three-body coupling is 
a quadratic polynomial in the coupling analogous to Eq.~\eqref{model_running}. 
The behavior of its zeros as functions of $d$ is compatible with the mechanism
for the loss of conformality in Ref.~\cite{Kaplan:2009}.

\section{Two-Body Sector}
\label{two_body}

Nonrelativistic particles with short-range interactions in three spatial dimensions have a nontrivial {\it zero-range limit} 
in which the range of the interaction is taken to zero with the s-wave scattering length $a$ held fixed. 
In the zero-range limit, $a$ is the only interaction parameter, at least in the two-body sector.
The {\it unitary limit} is defined by $a\rightarrow\pm\infty$. In the unitary limit, 
the two-body system is scale invariant, because the interactions no longer provide any length scale.  
The two-body scattering cross section has power-law dependence on  the collision energy $E$:
it saturates the s-wave unitarity bound $8\pi/E$.

We now consider identical bosons with zero-range interactions in $d$ dimensions. 
The T-matrix element ${\cal T}(k)$ for the scattering of two 
particles with total energy $E=k^2$ in the center-of-momentum frame can be 
derived by solving a Lippmann-Schwinger equation:
\begin{equation}
{\cal T}(k)=\frac{2(4\pi)^{d/2}/\Gamma\big(\frac{2-d}{2}\big)}{1/\lambda- e^{-i\pi (d-2)/2} k^{d-2}}.
\label{T-matrix}
\end{equation}
For simplicity, we set $\hbar=1$ and $m=1$ here and in the remainder of this article. 
The single interaction parameter $\lambda$ that characterizes the zero-range limit 
is proportional to the T-matrix element at $k=0$. 
In $d=3$, the interaction parameter is the scattering length: $\lambda = a$.
The unitary limit is defined by $\lambda\rightarrow\pm\infty$.  In this limit, 
the T-matrix element in Eq.~\eqref{T-matrix} 
has power-law dependence on the momentum: ${\cal T}(k)\sim1/ k^{d-2}$.
Since the interactions no longer provide any length scale, the two-body system is scale invariant.
The T-matrix element in Eq.~\eqref{T-matrix} goes to 0 as $d \to 2$ and as $d \to 4$, 
indicating that  2 and 4 are critical dimensions 
where the theory becomes noninteracting.
The upper critical dimension $d=4$ was first pointed out by Nussinov and Nussinov \cite{Nussinov:2004}.

Identical bosons in the zero-range limit can be described by a local quantum field theory with an atom field $\psi$. The 
Lagrangian density is
\begin{equation}
 {\cal L}=\psi^{\dagger}\left(i\frac{\partial}{\partial t}+\frac{1}{2}\nabla^2\right)\psi
-\frac{g_2}{4}\left(\psi^{\dagger}\psi\right)^2-\frac{g_3}{36}\left(\psi^{\dagger}\psi\right)^3,
\label{Lagrangian_two_body}
\end{equation}
where $g_2$ and $g_3$ are bare couplings for 
the two-body and three-body contact interaction. The 
two-body coupling $g_2$ can be tuned as a function of the UV cutoff to get the desired interaction
parameter $\lambda$. In three dimensions, the three-body coupling $g_3$ is required to reproduce 
Efimov physics \cite{Bedaque:1999, Hammer:1999}. No higher-body couplings are required.
We will  determine the RG flow of $g_2$  in $d$ dimensions and use it to understand scale invariance in the 
two-body sector in the unitary limit.

The perturbative expansion of the off-shell two-body scattering amplitude in powers of $g_2$ is UV divergent.
The amplitude can be regularized by imposing a sharp UV cutoff $\Lambda$ on loop momenta
in the center-of-mass frame.\footnote{
The same expression for the two-body amplitude can be obtained 
by using dimensional regularization with power divergence subtraction \cite{Kaplan:1998tg}
and with renormalization scale $\Lambda/2$, but no one has succeeded in 
deriving Efimov physics in the 3-body sector using dimensional regularization.} 
The bare coupling $g_2$ must depend on $\Lambda$ 
to compensate for the explicit dependence of the two-body scattering amplitude on $\Lambda$.
In order to reproduce the T-matrix element in Eq.~\eqref{T-matrix}, 
$g_2$ must be related to the interaction parameter $\lambda$ via
\begin{equation}
g_2(\Lambda)=-\frac{2(4\pi)^{d/2}}{\Gamma\big(\frac{2-d}{2}\big)}
\left[\frac{1}{\lambda}+\frac{2\sin\!\big(\frac{d}{2}\pi\big)}{(d-2)\pi}\Lambda^{d-2}\right]^{-1}.
\label{renormalized_coupling}
\end{equation}
We define a dimensionless two-body coupling $\hat{g}_2$ by multiplying 
 $g_2$ by the power of the UV cutoff $\Lambda$ required by dimensional analysis:
\begin{equation}
\hat{g}_2(\Lambda)=\frac{1}{(d-2)(4\pi)^{d/2}
	\Gamma\big(\frac{d}{2}\big)} \Lambda^{d-2} g_2(\Lambda).
\label{dimensionless_g2}
\end{equation}
The  $d$-dependent prefactor has been introduced in order to simplify the RG equation 
for $\hat{g}_2(\Lambda)$ below.
The RG equation for $\hat{g}_2$ can be obtained by differentiating 
both sides of Eq.~\eqref{dimensionless_g2} with respect to  $\log \Lambda$
and then eliminating $\Lambda$ in favor of $\hat{g}_2(\Lambda)$: 
\begin{equation}
\Lambda\frac{d~}{d\Lambda} \hat{g}_2=
(d-2)\hspace{0.1 cm}\hat{g}_2\left(\hat{g}_2 + 1\right).
\label{differential_RG}
\end{equation}

Scale invariance arises at a fixed point of the RG flow  for $\hat{g}_2$ or, equivalently, 
at a zero of the beta function defined by the right side of 
Eq.~\eqref{differential_RG}.
The RG flow for $\hat{g}_2$ has two fixed points:  
a UV fixed point $\hat{g}_{\rm +}=-1$ and an IR fixed point $\hat{g}_{\rm -}=0$. 
Using  Eqs.~\eqref{renormalized_coupling}  and \eqref{dimensionless_g2},
we find that the UV fixed point corresponds to the
unitary limit $\lambda\rightarrow\pm\infty$ and the IR fixed point corresponds to the
non-interacting limit  $\lambda \to 0$.
At the UV fixed point, there are scale-invariant interactions in the two-body sector.

\section{Three-body Sector}\label{three_body_Efimov}

In 1970,  Vitaly Efimov  discovered that a system of three identical bosons 
 in three spatial dimensions in the unitary limit $a\rightarrow\pm\infty$
 has a sequence of infinitely many shallow three-body bound states 
(Efimov states) whose binding energies have an accumulation point at the energy of the three-body scattering 
threshold  \cite{Efimov:1970}. 
As they approach the threshold,
the ratio of the binding energies of successive  Efimov states approaches a 
universal factor that is approximately $515 \approx 22.7^2$.
The spectrum of Efimov states reflects a discrete scaling symmetry 
with discrete scaling factor $e^{\pi/s_0} \approx22.7$, where $s_0 =1.00624$.
In subsequent papers, Efimov showed that the system of three identical bosons has universal properties 
characterized by discrete scaling behavior with the discrete scaling factor $e^{\pi/s_0}$
not only in the unitary limit but whenever the scattering length $a$ 
is much larger than the range  of the interactions \cite{Efimov:1971, Efimov:1979}.

From an RG perspective, discrete scaling behavior should be associated with 
an RG flow whose ultraviolet attractor is a limit cycle. Bedaque, Hammer, and van Kolck 
showed that in $d=3$, the renormalization of the three-body 
coupling $g_3$ in Eq.~\eqref{Lagrangian_two_body} is indeed governed by a limit 
cycle \cite{Bedaque:1999, Hammer:1999}. Braaten and Hammer derived 
the RG equation for the dimensionless coupling $\hat{g}_3=\Lambda^4g_3/144\pi^4$ \cite{Eric_Hammer}:
\begin{equation}
\Lambda\frac{d~}{d\Lambda} \hat{g}_3
=\frac{1+s_0^2}{2}\left(\hat{g}_2^2+\frac{\hat{g}_3^2}{\hat{g}_2^2}\right)+\left(3-s_0^2+2\hat{g}_2\right)\hat{g}_3.
\label{RG_equation_d_3}
\end{equation}
The beta function defined by the right side is a quadratic polynomial in $\hat{g}_3$
with coefficients that depend on $\hat g_2$.
This structure is reminiscent of the beta function in Eq.~\eqref{model_running} that describes the loss of conformality. 
Near the UV fixed point $\hat{g}_2=-1$ of the two-body coupling, 
the zeros of the beta function in Eq.~\eqref{RG_equation_d_3} are complex.  
Thus the RG flow  for $\hat{g}_2$ and $\hat{g}_3$ does not have a  UV fixed point.
Instead the coupling constants flow in the ultraviolet to a limit cycle given by
\begin{equation}
\hat{g}_2(\Lambda)=-1, \quad \hat{g}_3(\Lambda)=-\frac{\cos\left[s_0\log\left(\Lambda/\Lambda_*\right)
	+\arctan s_0\right]}{\cos\left[s_0\log\left(\Lambda/\Lambda_*\right)
	-\arctan s_0\right]},
\label{g2g3_limitcycle}
\end{equation}
where $\Lambda_*$ is a constant momentum scale.

\subsection{Atom-diatom amplitude}

In Ref.~\cite{Bedaque:1999, Hammer:1999}, Bedaque, Hammer, and van Kolck [BHvK] 
developed an effective field theory (EFT) to calculate universal results 
for three-body observables for three identical bosons in the zero-range limit  in three spatial dimensions. 
We will use this EFT to describe the system in a variable spatial dimension $d$.
The fields in this EFT are the dynamical atom field $\psi$ and an auxiliary diatom field $\Delta$. 
The BHvK Lagrangian density is 
\begin{equation}
 {\cal L}_{\rm BHvK}=\psi^{\dagger}\left(i\frac{\partial}{\partial t}+\frac{1}{2}\nabla^2\right)\psi+\frac{g_2}{4}\Delta^{\dagger}\Delta-\frac{g_2}{4}\left(
\Delta^{\dagger}\psi^2+\psi^{\dagger 2}\Delta\right)-\frac{g_3}{36}\left(\Delta\psi\right)^{\dagger}\left(\Delta\psi\right),
\label{Lagrangian_three_body}
\end{equation}
where $g_2$ and $g_3$ are bare two-body and three-body couplings. 
Using the equation of motion for $\Delta$, 
we can integrate out the diatom field $\Delta$ analytically to obtain the 
 Lagrangian density 
\begin{equation}
{\cal L}=\psi^{\dagger}\left(i\frac{\partial}{\partial t}+\frac{1}{2}\nabla^2\right)\psi-\frac{g_2}{4}\frac{\left(\psi^{\dagger}\psi\right)^2}{1-\left(g_3/9g_2\right)\psi^{\dagger}\psi}.
\label{Lagrangian_integrating_delta}
\end{equation}  
Expanding the interaction term  in powers of $\psi^{\dagger}\psi$ and truncating after the
$\left(\psi^{\dagger}\psi\right)^3$ term, we recover the Lagrangian for 
identical bosons in Eq.~\eqref{Lagrangian_two_body}. The additional operators  
$\left(\psi^{\dagger}\psi\right)^n$ with $n\ge 4$ from the 
expansion of the interaction potential in Eq.~\eqref{Lagrangian_integrating_delta} are 
irrelevant operators at the UV fixed point $\hat g_2 = -1$ for the two-body coupling,
so they can  be ignored. Therefore, the quantum field theory 
described by the BHvK Lagrangian density in Eq.~\eqref{Lagrangian_three_body} 
is equivalent to the quantum field theory described by the 
Lagrangian density in Eq.~\eqref{Lagrangian_two_body}.  

\begin{figure}[t]
	\centering
	\centerline{ \includegraphics[width=9 cm,clip=true]{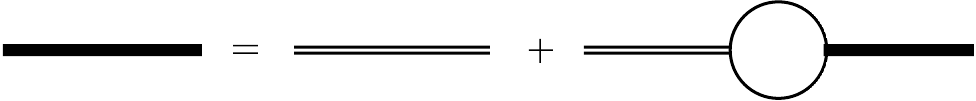} }
	\vspace*{0.0cm}
	\caption{Diagrammatic equation for the complete diatom propagator $iD$.}
	\label{quantum_corrections}
\end{figure}

The propagator of the auxiliary diatom field $\Delta$ is $4i/g_2$. It receives quantum corrections 
from a geometric series  of bubble diagrams 
that can be obtained by iterating the diagrammatic equation in Fig.~\ref{quantum_corrections}.
The complete diatom propagator in $d$ spatial dimensions including the quantum corrections is
\begin{equation}
iD\left(p_{\rm 0},{\bm p}\right)=-i\frac{8(4\pi)^{d/2}}{\Gamma\big(\frac{2-d}{2}\big)g_2(\Lambda)^2}
\left[ 1/\lambda -\left(-p_{\rm 0}+p^2/4-i\epsilon\right)^{(d-2)/2}\right]^{-1}.
\label{diatom_running}
\end{equation}
All the dependence on the UV cutoff $\Lambda$ is in the multiplicative factor $1/g_2^2$. 
In the unitary limit $\lambda \to \pm \infty$, the propagator in Eq.~\eqref{diatom_running}
has a pole in $p_0$ at the upper critical dimension $d=4$.

\begin{figure}[t]
\centering
	 \includegraphics[width=12 cm,clip=true]{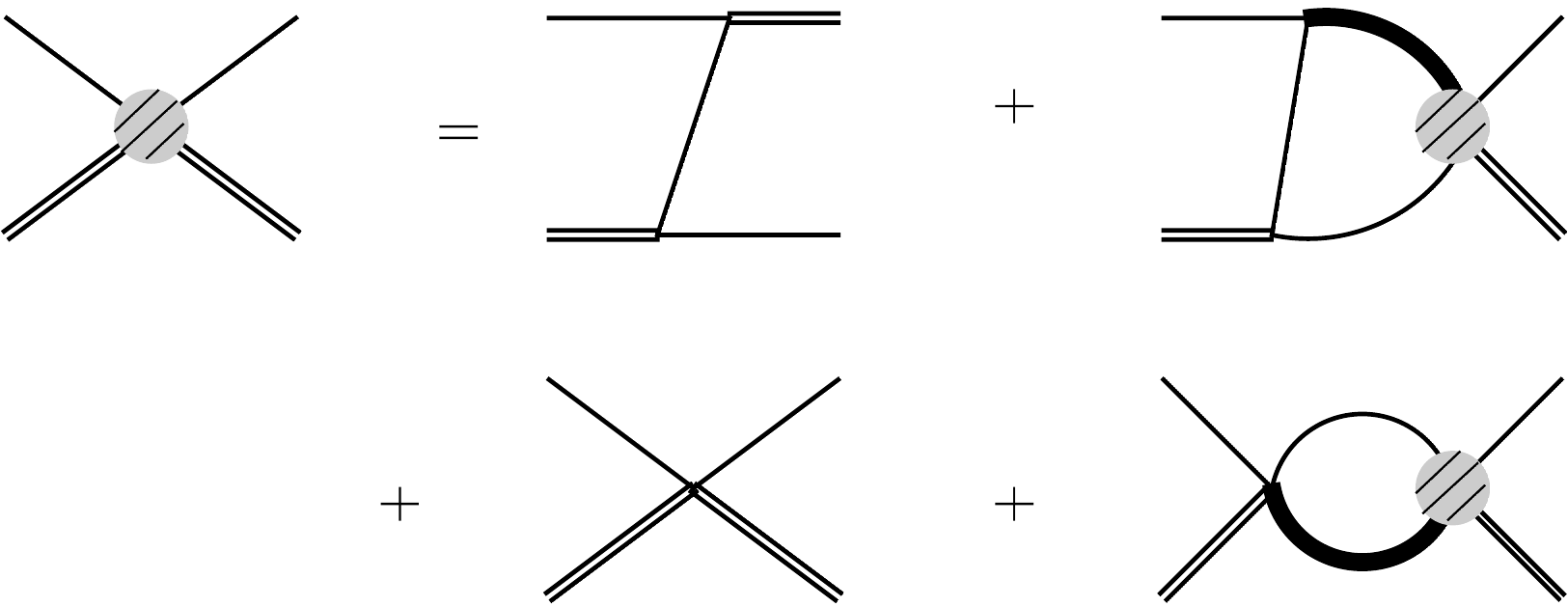} 
	\vspace*{0.0cm}
	\caption{Integral equation for the atom-diatom amplitude $\mathcal{A}$.
	}
	\label{diagrammatic_amplitude}
\end{figure}

The dependence of the three-body coupling $g_3$ on the UV cutoff $\Lambda$ can be determined from the off-shell
atom-diatom amplitude $\mathcal{A}$. The amplitude $\mathcal{A}$ satisfies the
 integral equation shown diagrammatically in Fig.~\ref{diagrammatic_amplitude}.
If $g_3=0$, the diagrams in the second  line of Fig.~\ref{diagrammatic_amplitude} are zero,
and the diagrammatic equation reduces to the Skorniakov-Ter-Martirosian (STM) integral equation \cite{STM:1957}.
In the limit $\Lambda \to \infty$, the solutions of the STM equation are not ultraviolet divergent, but they depend log-periodically on $\Lambda$.
Requiring the log-periodic dependence  to be cancelled by the diagrams in the second  line of Fig.~\ref{diagrammatic_amplitude}
determines the dependence of $g_3$ on  $\Lambda$.

To obtain the RG equation for the three-body coupling $g_3$, 
we follow closely the path used in Ref.~\cite{Eric_Hammer} to derive the 
beta function in Eq.~\eqref{RG_equation_d_3}, except that we carry out all the steps in dimension $d$ instead 
of $d=3$. In the center-of-mass frame, the atom-diatom amplitude is a function of the
incoming relative momentum ${\bm p}$, the outgoing relative momentum ${\bm k}$, and the total energy $E$. 
The s-wave atom-diatom amplitude $\mathcal{A}_s(p,k;E)$ can be obtained by averaging 
$\mathcal{A}({\bm p},{\bm k};E)$ over the angle $\theta$
between ${\bm p}$ and ${\bm k}$ in $d$ dimensions.
To simplify the integral equation satisfied by $\mathcal{A}_s$, we also choose to multiply it by $4/g_{\rm 2}^2$:
\begin{equation}
\mathcal{A}_s(p,k;E)=\frac{4}{g_{\rm 2}^2}
\frac{\Gamma\big(\frac{d}{2}\big)}{\Gamma\big(\frac{d-1}{2}\big)\Gamma\big(\frac{1}{2}\big)}
\int_{-1}^{+1}d\cos\theta\hspace{0.1 cm}\left(1-\cos^2\theta\right)^{\left(d-3\right)/2}\mathcal{A}({\bm p},{\bm k};E).
\label{angle_averaging}
\end{equation}
The integral equation for  $\mathcal{A}_s$
with a sharp UV cutoff $\Lambda$ on the loop momentum in the center-of-mass frame is
\begin{eqnarray}
\mathcal{A}_s\left(p,k;E\right)&=&-
\left[\frac{1}{E-p^2-k^2+i\epsilon}\, {_2}F_1\!\left({\textstyle\frac{1}{2},1\atop\textstyle\frac{d}{2}}\,
{\Bigg |}\frac{p^2k^2}{\left(E-p^2-k^2+i\epsilon\right)^2}
	\right)+\frac{\hat{G}(\Lambda)}{\Lambda^2}\right]\nonumber\\
&&-\frac{4\sin\big(\frac{d}{2}\pi\big)}{\pi}
\int_0^\Lambda dq\hspace{0.1 cm}q^{d-1} 
\left[\frac{1}{E-p^2-q^2+i\epsilon} \, {_2}F_1\!\left({\textstyle\frac{1}{2},1\atop\textstyle\frac{d}{2}}\,
{\Bigg |}\frac{p^2q^2}{\left(E-p^2-q^2+i\epsilon\right)^2}\right)\right.\nonumber\\
&&\hspace{5.0 cm}\left.+\frac{\hat{G}(\Lambda)}{\Lambda^2}\right]
\frac{\mathcal{A}_s\left(q,k;E\right)}{1/\lambda-
	\left(-E+ 3q^2/4-i\epsilon\right)^{(d-2)/2}},
\label{integral_equation}
\end{eqnarray}
where $\hat{G}(\Lambda)$ is a dimensionless three-body coupling defined by
\begin{equation}
\hat{G}(\Lambda)=\frac{\Lambda^2 g_3(\Lambda)}{9g_2(\Lambda)^2} .
\label{dimensionless_H}
\end{equation}
The dependence of the amplitude $\mathcal{A}_s$ on $\Lambda$ has been suppressed in Eq.~\eqref{integral_equation}.
The dependence of $\hat{G}$ on $\Lambda$ can be determined by requiring the solutions to 
Eq.~\eqref{integral_equation}  to have well-behaved limits as $\Lambda \to \infty$ \cite{Bedaque:1999,Hammer:1999}.
Since $\hat{G}$ is the only coupling in the integral equation,
the beta function for $\hat{G}$ can depend only on $\hat{G}$.

\subsection{Asymptotic solutions}

In the limit $\Lambda \to \infty$, the integral equation in  Eq.~\eqref{integral_equation} for the s-wave 
atom-diatom amplitude 
$\mathcal{A}_s(p,k,E)$ has asymptotic solutions for large $p$ that have  power-law dependence $p^{s-1}$.
The possible  exponents of $p$ can be determined by neglecting the inhomogeneous  terms in Eq.~\eqref{integral_equation}, 
replacing $\mathcal{A}_s(p,k,E)$ by $A p^{s-1}$,
setting $E$, $\hat{G/\Lambda^2}$, and $1/\lambda$
to 0 inside the integral over $q$, and taking the upper endpoint $\Lambda$ of the integral to $\infty$  \cite{Eric_Hammer}.
After making the change of variable $q=xp$, the dependence on $p$ drops out and 
the integral equation in Eq.~\eqref{integral_equation} reduces to
\begin{equation}
1=-\left(\frac{4}{3}\right)^{\frac{d-2}{2}}\frac{4\sin\big(\frac{d}{2} \pi \big)}{\pi}
\int_0^{\infty}dx\hspace{0.1cm}\frac{x^{s}}{1+x^2}\hspace{0.1 cm}
_2F_1\! \left({\frac{1}{2},1\atop \frac{d}{2}}\,{\Bigg |}\frac{x^2}{\left(1+x^2\right)^2}
	\right).
\label{scaling_equation_1}
\end{equation}
The integral can be evaluated analytically by inserting the power 
series definition of the hypergeometric function $_2F_1$. After integrating 
the individual terms of the power series over $x$ and then resumming the series, the resulting 
equation for $s$ is 
\begin{equation}
2\sin\left(\frac{d}{2}\pi\right){_2}F_1\!
\left({\frac{d-1+s}{2}, \frac{d-1-s}{2} \atop  \frac{d}{2}}\,{\Bigg |}\frac{1}{4}\right)+\cos\left(\frac{ s}{2}\pi\right)=0.
\label{scaling_solution}
\end{equation}
The equation is invariant under $s\rightarrow -s$, so it is an equation for $s^2$. 

\begin{figure}[t]
	\includegraphics[width=10cm,clip=true]{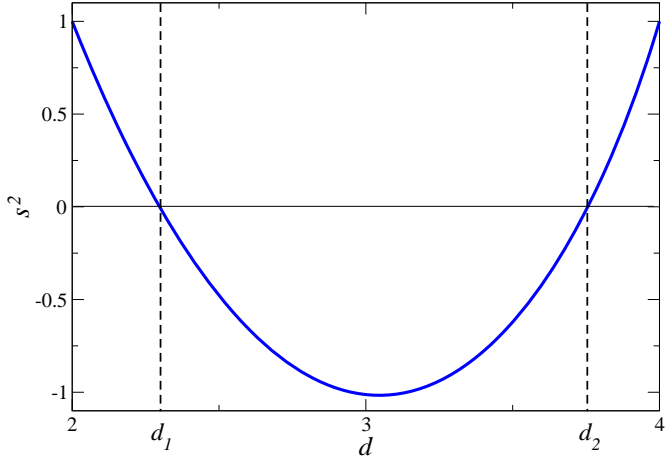} 
	\vspace*{0.0cm}
	\caption{The lowest branch of solutions of Eq.~\eqref{scaling_solution} for 
	$s^2$ as a function of the dimension $d$. 
	The solution for $s^2$ is negative in the interval $d_{\rm 1}<d<d_{\rm 2}$ 
	between the two vertical dotted lines.}
	\label{svsd}
\end{figure}

The equation in Eq.~\eqref{scaling_solution}
has infinitely many branches of solutions for $s^2$ as a function of the dimension $d$.
The lowest branch of solutions is shown in Fig.~\ref{svsd}. 
The value of $s^2$ decreases from 1 at the endpoints $d=2, 4$ to a minimum of $-1.016$ at $d=3.04$.
The higher branches of solutions have  $s^2>9$.  
The only branch that is physically relevant is the lowest branch.
The higher branches are irrelevant, because 
the integral over $x$ in Eq.~\eqref{scaling_equation_1} is convergent at both the endpoints $0$ and $\infty$ 
only if the exponent $s$ satisfies the condition $-1<{\rm Re}\left(s\right)<1$. 
There are two critical dimensions at which Eq.~\eqref{scaling_solution} is 
satisfied by $s^2=0$: 
 \begin{equation}
d_{\rm 1}=2.30,\quad d_{\rm 2}=3.76.
\label{solution_critical_dimension}
\end{equation}
They are the lower and upper critical dimensions for the Efimov effect \cite{Nielsen:2001}.
We will focus first on the 
regions $2<d<d_{\rm 1}$ and $d_{\rm 2}<d<4$ and then on the region $d_{\rm 1}<d<d_{\rm 2}$.

\subsection{Conformality}
{\label{RG_fixedpoint}}

In the regions $2<d<d_{\rm 1}$ and $d_{\rm 2}<d<4$, 
the lowest branch of solutions of Eq.~\eqref{scaling_solution} for the exponent of $p$ 
 are  $\pm s-1$,
where $s$ is real and positive.
The value of $s$ depends on $d$, decreasing from 1 to 0 as $d$ goes from 2 to $d_1$
and as $d$ goes from 4 to $d_2$.
The most general asymptotic solution 
of the integral equation in Eq.~\eqref{integral_equation} as $p\rightarrow\infty$ is given by
\begin{equation}
\mathcal{A}_s\left(p,k;E\right)\longrightarrow A_{+}p^{s-1}+ A_{-}p^{-s-1},
\label{asymptotic_scaling_solution}
\end{equation}
where $A_{\pm}$ are two constants that may depend on $k$, $E$, and $\lambda$.
The inhomogeneous term in the integral equation
determines one of the constants and
the three-body coupling determines the other constant. 

The $\Lambda$ dependence of the dimensionless coupling $\hat{G}(\Lambda)$ can be obtained by  
inserting the asymptotic solution in Eq.~\eqref{asymptotic_scaling_solution} into 
the integral equation in Eq.~\eqref{integral_equation}
 and requiring the integral over $q$ to be independent of $\Lambda$ \cite{Eric_Hammer}.
The $\Lambda$ dependence of the integral  is proportional to
\begin{eqnarray}
&&\int^{\Lambda}dq\,\left(\frac{1}{q^2}-\frac{\hat{G}(\Lambda)}{\Lambda^2}\right) \left(A_{+}q^{s}+A_{-}q^{-s}\right)
\nonumber
\\ 
&& \hspace{3.5cm}
= \left( \frac{A_{+}\Lambda^{s-1}}{s-1}+ \frac{A_{-}\Lambda^{-s-1}}{-s-1} \right) 
-  \hat{G}(\Lambda) \left( \frac{A_{+}\Lambda^{s-1}}{s+1}+\frac{A_{-}\Lambda^{-s-1}}{-s+1} \right).
\label{determining_H_fixedpoint}
\end{eqnarray}
Setting this equal to 0 gives an
expression for $\hat{G}(\Lambda)$ that depends on the ratio $A_{+}/A_{-}$. 
It can be simplified by
setting $A_{+}/A_{-} = \Lambda^{-2s}_{*}$, where $\Lambda_*$ is a constant 
momentum scale. The resulting expression for $\hat{G}(\Lambda)$ is
  \begin{equation}
\hat{G}(\Lambda)=-\frac{\cosh\left[s\log\left(\Lambda/\Lambda_*\right)
	+\arctanh s\right]}{\cosh\left[s\log\left(\Lambda/\Lambda_*\right)
	-\arctanh s\right]}.
\label{RG_variable_fixedpoint}
 \end{equation}
The RG equation for $\hat{G}(\Lambda)$ can be obtained by
differentiating both sides of Eq.~\eqref{RG_variable_fixedpoint} with respect to $\log \Lambda$
and then eliminating $\Lambda$ in favor of $\hat{G}(\Lambda)$: 
\begin{equation}
 \Lambda\frac{d~}{d\Lambda} \hat{G}=\frac{1-s^2}{2}+\left(1+s^2\right)\hat{G}
 +\frac{1-s^2}{2}\hat{G}^2,
 \label{RG_three_fixedpoint}
 \end{equation}
  where $s^2$ is the function of $d$ defined by Eq.~\eqref{scaling_solution}.
 The beta function defined by the right side of Eq.~\eqref{RG_three_fixedpoint} is a quadratic polynomial in $\hat{G}$. 
 The RG flow for $\hat{G}$ has two fixed points where the beta function is zero: 
 a UV fixed point $\hat{G}_{\rm +}$ and an IR fixed point $\hat{G}_{\rm -}$  given by 
 \begin{equation}
\hat{G}_{\rm \pm}=-\frac{1\pm s}{1\mp s}.
 \label{three_body_fixedpoints}
 \end{equation}
As $d$ increases towards $d_{\rm 1}$ or as $d$ decreases towards $d_{\rm 2}$, 
$s$ approaches 0 so the two fixed points $\hat{G}_{\rm -}$ and $\hat{G}_{\rm +}$ both approach  $-1$.  
At unitarity, the two-body coupling is at its UV fixed point  $\hat{g}_2=-1$. If $\hat{G}$ is also at its UV fixed point $G_+$,
the three-body sector is scale invariant and therefore presumably also conformally invariant.

\subsection{Conformality lost}

In the region $d_{\rm 1}<d<d_{\rm 2}$, 
the lowest branch of solutions of Eq.~\eqref{scaling_solution} for the exponent of $p$ are  $\pm i s_0-1$,
where $s_0$ is real and positive.
The value of $s_0$ depends on $d$, increasing from 0 to 1.00812 and then decreasing to 0 as $d$ increases from 
$d_1$ to 3.04 and then to $d_2$.   At $d=3$, its value is $s_0 =1.00624$. 
The most general asymptotic solution 
of the integral equation in Eq.~\eqref{integral_equation}  as $p\rightarrow\infty$ is given by 
\begin{equation}
\mathcal{A}_s\left(p,k;E\right)\longrightarrow A_{+}p^{is_0-1}+A_{-}p^{-is_0-1},
\label{asymptotic_scaling_limitcycle}
\end{equation}
where $A_{\pm}$ are two constants that may depend on $k$, $E$, and $\lambda$.

Using the arguments in Section~\ref{RG_fixedpoint}, one can again determine 
the $\Lambda$ dependence of the dimensionless parameter $\hat{G}(\Lambda)$. 
Upon inserting the asymptotic solution in Eq.~\eqref{asymptotic_scaling_limitcycle} into
the integral equation in Eq.~\eqref{integral_equation}, the dependence of the integral on $\Lambda$
has the form  in Eq.~\eqref{determining_H_fixedpoint} with $s$ replaced by $i s_0$.
The dependence on $\Lambda$ cancels if $\hat{G}(\Lambda)$ has the form
\begin{equation}
\hat{G}(\Lambda)=-\frac{\cos\left[s_0\log\left(\Lambda/\Lambda_*\right)
	+\arctan s_0\right]}{\cos\left[s_0\log\left(\Lambda/\Lambda_*\right)
	-\arctan s_0\right]},
\label{RGvariable_limitcycle}
\end{equation}
where $\Lambda_*$ is a constant momentum scale.
This expression is a log-periodic function of $\Lambda$ with period $e^{\pi/s_0}$, so the RG flow of $\hat{G}$ is a limit cycle. 
The RG equation for $\hat{G}$ can 
be obtained by differentiating both sides
of Eq.~\eqref{RGvariable_limitcycle} with respect to $\log \Lambda$
and then eliminating $\Lambda$ in favor of $\hat{G}(\Lambda)$.
The result is just the RG equation for $\hat{G}$ in Eq.~\eqref{RG_three_fixedpoint} with $s^2=-s_0^2$.
Since $s^2<0$, the zeros of this beta function are complex valued.
At unitarity, the two-body coupling is at its UV fixed point  $\hat{g}_2=-1$,
so the two-body sector is scale invariant. However the three-body sector is not scale invariant.
Instead it has discrete scale invariance with discrete scaling factor $e^{\pi/s_0}$.

We can define a dimensionless three-body coupling $\hat{g}_{\rm 3}$ by multiplying $g_{\rm 3}$ with the power of 
UV cutoff $\Lambda$ required by dimensional analysis:
\begin{equation}
\hat{g}_{\rm 3}\left(\Lambda\right)=\frac{1}{\left[3(d-2)(4\pi)^{d/2}\Gamma\big(\frac{d}{2}\big)\right]^2}
\Lambda^{2d-2}g_{\rm 3}\left(\Lambda\right).
\label{dimensionless-g3}
\end{equation}
The $d$-dependent prefactor has been introduced in order to simplify the RG equation for $\hat{g}_{\rm 3}$ below.
The RG equation for  $\hat{g}_3 =\hat{g}_2^2 \, \hat{G} $
can be derived by using the RG equations for $\hat{G}(\Lambda)$ in Eq.~\eqref{RG_three_fixedpoint} 
and for
$\hat{g}_2$ in Eq.~\eqref{differential_RG}: 
\begin{equation}
\Lambda\frac{d~}{d\Lambda} \hat{g}_3
=\frac{1-s^2}{2}\left(\hat{g}_2^2+\frac{\hat{g}_3^2}{\hat{g}_2^2}\right)+\left(2d-3+s^2+2(d-2)\hat{g}_2\right)\hat{g}_3,
\label{RG_equation_d}
\end{equation} 
where $s^2$ is the function of $d$ defined by Eq.~\eqref{scaling_solution}.
Upon setting $d=3$ and $s^2 = - s_0^2$, where $s_0 = 1.00624$,
this reduces to the RG equation in Eq.~\eqref{RG_equation_d_3} derived in Ref.~\cite{Eric_Hammer}.

\section{Summary}\label{Summary}

In Ref.~\cite{Kaplan:2009}, Kaplan, Lee, Son, and  Stephanov revealed a general mechanism for the loss of 
conformal invariance in a system. Near the point where conformality is lost, the beta function of an 
appropriate coupling is a quadratic function of the coupling as in 
Eq.~\eqref{model_running}. As an external parameter is tuned to the critical value 
where conformality is lost,  an IR fixed point and a UV fixed point of the coupling merge together 
and then disappear.
 
In this work, we have shown explicitly the mechanism for loss of conformal invariance 
 associated with Efimov physics in the case of 
identical bosons at unitarity with variable spatial dimension $d$. The beta 
function for the dimensionless  three-body coupling $\hat{G}$ defined in Eq.~\eqref{dimensionless_H}
is the quadratic polynomial in $\hat{G}$ 
given in Eq.~\eqref{RG_three_fixedpoint}. Its coefficients depend on $d$
through $s^2$, which is a solution to Eq.~\eqref{scaling_solution}. There are two critical 
dimensions $d_{\rm 1}=2.30$ and $d_{\rm 2}=3.76$ at which conformality is lost. 
In the regions $2 < d < d_1$ and $d_2 < d < 4$, the beta function for  
$\hat{G}$ has real zeros and its RG flow has IR and UV fixed points. 
As $d$ increases through $d_{\rm 1}$ or decreases through
$d_{\rm 2}$, the IR and UV fixed points approach each other, merge together, and then disappear into the complex plane. 
In the region $d_{\rm 1}<d<d_{\rm 2}$ where  
conformality is lost, the beta function for $\hat{G}$ 
has complex zeros and its RG flow is a limit cycle. 

In the case of two species of fermions in 3 spatial dimensions, 
the ratio $M/m$ of the masses plays the role of an external parameter 
that controls the loss of conformality.
The Efimov effect occurs in the $p$-wave channel at unitarity for two heavy fermions and one light fermion 
when the mass ratio $M/m$ exceeds 13.6 \cite{Efimov:1973}. 
It would be interesting to  provide an RG perspective on
the loss of conformality associated with Efimov physics in  this
system by determining the beta function for an appropriate 3-body coupling.

The unitary Fermi gas, which consists of fermions with two spin states and infinitely large scattering length
in 3 spatial dimensions, is a challenging many-body physics problem.  
The most successful analytic approach to this problem 
has been through interpolation in the number of spatial dimensions $d$ using epsilon expansions 
around the critical dimensions $d=2$ and $d=4$ \cite{Nishida:2006eu}.
The ground state energy for the unitary Fermi gas  has been calculated using next-to-next-to-leading-order 
epsilon expansions around both 2 and 4 \cite{Nishida:2006eu,Arnold:2006fr,Nishida:2008mh}.
Interpolation in $d$ is essential because of the poor convergence properties of the epsilon expansions around 2 and 4.
The unitary Bose gas, which consists of identical bosons with infinitely large scattering length
in 3 spatial dimensions, is an even more challenging many-body physics problem because of the Efimov effect.  
The epsilon expansion around the upper critical dimension 4 has been applied to the unitary Bose gas \cite{JLSZ:2014}.
However interpolation in $d$ using epsilon expansions around both 2 and 4 seems to be more promising.
The understanding of the RG flow near the critical dimensions
2.30 and 3.76 for the loss of conformality that we have presented in this paper
could be essential for applying 
interpolation in $d$ to the unitary Bose gas in 3 dimensions.

\begin{acknowledgments}
	This research was supported by the National Science Foundation under grant PHY-1607190.
\end{acknowledgments}

\end{document}